\documentclass{WileyMSP-template}
\usepackage{amssymb,amsmath,mathrsfs,mathtools,bbm,bbold}
\usepackage[usenames, dvipsnames]{xcolor}
\usepackage{graphicx}
\usepackage{notes2bib}

\let\footnote=\bibnote

\newcommand{\cir}{\mathscr{C}}

\newcommand{\cE}{\mathcal{E}}

\newcommand{\cN}{\mathcal{N}}
\newcommand{\cF}{\mathcal{F}}

\newcommand{\cP}{\mathcal{P}}

\newcommand{\sM}{\mathscr{M}}
\newcommand{\sR}{\mathscr{R}}
\newcommand{\sG}{\mathscr{G}}

\newcommand{\id}{\mathbb{1}}
\newcommand{\PL}{P_\mathrm{UB}(L)}
\newcommand{\WL}{W_\mathrm{UB}(L)}
\newcommand{\WpL}{W'_\mathrm{UB}(L)}

\begin{document}

\title{On the Fault-Tolerance Threshold for Surface Codes with General Noise}

\maketitle

\author{Jing Hao Chai* and Hui Khoon Ng*}

\bigskip
\begin{affiliations}
Dr.~J.~H.~Chai\\
Centre for Quantum Technologies, National University of Singapore,\\
3 Science Drive 2, Singapore 117543, Singapore\\
Email Address: jh.chai@u.nus.edu\\
\medskip
Assoc.~Prof.~H.~K.~Ng\\
Yale-NUS College, 16 College Avenue West, Singapore 138527\\
Centre for Quantum Technologies, National University of Singapore\\
Department of Physics, National University of Singapore\\
MajuLab, CNRS-UNS-NUS-NTU International Joint Unit, UMI 3654, Singapore\\
Email Address: huikhoon.ng@yale-nus.edu.sg

\end{affiliations}


\begin{abstract}
Fault-tolerant quantum computing based on surface codes has emerged as a popular route to large-scale quantum computers capable of accurate computation even in the presence of noise. Its popularity is, in part, because the fault-tolerance or accuracy threshold for surface codes is believed to be less stringent than competing schemes. This threshold is the noise level below which computational accuracy can be increased by increasing physical resources for noise removal, and is an important engineering target for realising quantum devices. The current conclusions about surface code thresholds are, however, drawn largely from studies of probabilistic noise. While a natural assumption, current devices experience noise beyond such a model, raising the question of whether conventional statements about the thresholds apply. Here, we attempt to extend past proof techniques to derive the fault-tolerance threshold for surface codes subjected to general noise with no particular structure. Surprisingly, we found no nontrivial threshold, i.e., there is no guarantee the surface code prescription works for general noise. While this is not a proof that the scheme fails, we argue that current proof techniques are likely unable to provide an answer. A genuinely new idea is needed, to reaffirm the feasibility of surface code quantum computing. 
\end{abstract}

\section{Introduction}
The subject of quantum computing has seen an immense growth in interest over the past few years, with academic groups and industry partners keenly pursuing the realisation of small-scale devices and rapidly expanding the variety of problems such devices can tackle. The eventual goal of quantum computing, however, is large-scale computers capable of reliable computation for problem sizes large enough to genuinely exploit the advantage of a quantum approach over classical computers. To tackle large problems with sufficient accuracy to be useful, quantum computational circuits, built from physical components that are unavoidably noisy, have to be implemented in a manner robust against noise and the computational errors that can result. The subject of fault-tolerant quantum computing is about how one can compute more accurately, while using noisy components, by investing more physical resources to deal with the errors that arise. 
\\\medskip

Fault-tolerant quantum computing schemes are based on the technique of quantum error correction. By encoding computational information in a well-chosen part---the code space---of the physical quantum state space, quantum error correction allows errors in the encoded information to be detected, diagnosed, and corrected, provided few enough errors occurred to not exceed the code capability. However, the quantum error correction process, namely the syndrome measurement (for error detection and diagnosis) and the recovery (for correction), is itself carried out by physical components that are also noisy and error-prone. Furthermore, the use of error correction requires a typically significant increase in the number of physical components---more physical quantum registers (e.g., qubits) and physical operations (gates and measurements)---in the quantum computer, hence increasing the number of ways things can go wrong in the presence of noise. Error correction can thus actually cause a net increase, rather than decrease, in errors in the quantum computer. Fault-tolerant quantum computing is precisely about how to design the error-corrected quantum computing circuit in a manner than ensures a net removal of errors. Provided the strength of the physical noise is below a threshold level, a fault-tolerant quantum computing prescription enables us to effectively remove computational errors even with noisy components, and consequently increase computational accuracy, 
\\\medskip

This noise threshold is discussed in fault-tolerance literature as the quantum accuracy threshold (see, for example, the classic fault tolerance papers \cite{KnillLaflammeZurek,PreskillReliableQC,Ben-Or}). Specifically, the accuracy threshold (or fault-tolerance threshold) refers to a threshold level of noise below which a fault-tolerant quantum computing prescription---which tells us how to build quantum circuits from noisy components---is able to arbitrarily increase computational accuracy by increasing the error correction capability of the code, accompanied by an increase in physical resources needed for the error correction procedure. The accuracy threshold determines the point where the noise is weak enough for scalable quantum computing to be possible, and is hence an important engineering target for building quantum computing devices. Many past works on the theory of fault-tolerant quantum computing have focused on deriving estimates of the accuracy threshold for different fault-tolerant schemes, based on different types of codes, under different models of noise (see, for example, References \cite{AGP,TerhalBurkard,AKP,AliferisLeakage,HKPreskill}).
\\\medskip

In this work, we take a closer look at the accuracy threshold for fault-tolerant quantum computing based on surface codes \cite{KitaevPlanar,Dennis,FreedmanMeyer}. The surface code (see, for example, References \cite{TerhalMemory,MartinisFowler} for an introduction to the subject) is a type of topological code that has emerged as a promising candidate for fault-tolerant quantum computing. 
In recent years, significant advances have been made in the implementation of surface codes as a means towards fault-tolerant quantum computing. These include experiments \cite{corcoles2015demonstration,takita2016demonstration} that demonstrated the basic steps of surface code stabilizer measurements on five noisy qubits, as well as rudimentary error correction experiments on quantum chip architectures with qubits arranged in a planar array \cite{andersen2020repeated,chen2021exponential,WallraffSurfaceCode,IonTrapColorCode, erhard2021entangling,marques2022logical,ChineseSurfaceCode}.
While experimental progress continues to hinge on both the scale of the quantum chips as well as the quality of the qubits therein, notable theoretical advances have also been made in recent years. These include proposals to manipulate the encoded information (see, for example, \cite{MartinisFowler,LatticeSurgery,PokingHoles,Yoder_SurfaceCode, Litinsky,VuillotLatticeSurgery, brown2020fault, Bartlett2020braiding, chamberland2022circuitLatticeSurgery}), as well as discussions of how universal quantum computing could be realistically achieved with surface codes (see, for example, \cite{campbell2017roads, lao2018mapping, chamberland2020topological, BombinLogicalFT, chamberland2022universal}).
\\\medskip

The most basic form of the surface code is implemented on a square lattice of physical qubits and has the primary experimental advantage (as do many topological codes) that the error correction steps can be done with just nearest-neighbor coupling between qubits. The quantum information is stored as a global property of the entire code lattice, making it tolerant to localized errors. Compared to some of the earlier fault-tolerant schemes based on concatenated codes, surface codes require comparatively fewer ancillary qubits and complicated ancilla states that can be difficult to prepare in practice \cite{Dennis}. Due to these features, the surface code has become a popular pathway towards large-scale quantum computers. Already, it has underpinned proposals for the development of physical quantum hardware architecture, with trapped ions \cite{TrappedIonArray, TrappedIonArray2}, quantum dots \cite{QDotsArray}, nitrogen-vacancy centers \cite{NVarray}, superconducting qubits \cite{Topoarray}, etc.\\
\medskip

Another reason for surface code's popularity is its seemingly more forgiving accuracy threshold, with less stringent requirements on the control of noise than many earlier fault-tolerant schemes. This conclusion comes from past work that can be classified into two main types of analyses \footnote{Reference \cite{BravyiCoherent} does not fall into these two categories. It studies code-capacity performance under coherent noise, using a novel analytical technique to allow for faster numerical simulation at larger code distances. The authors concluded that coherent noise is not detrimental to the performance of surface codes, when syndrome measurements are assumed to be ideal.}: (1) a phase transition argument \cite{Dennis,WangPreskill,NovaisCorrelated,ChubbsFlammia} that makes use of the topological nature of the code, and (2) analytical and numerical circuit-level studies for probabilistic noise \cite{Dennis,FowlerProof,StephensThreshold,MartinisFowler,BiasedNoiseThreshold_bad}. The former phase transition argument lacks the detailed circuit-level calculations needed to derive a rigorous accuracy threshold that fully accounts for the fault-tolerant error correction procedure, while the latter studies on probabilistic noise may not be applicable for actual quantum devices. In current devices, noise processes include, for example, spontaneous decay (modeled as amplitude-damping noise) as well as over- or under-rotation in gate operations due to misalignment (modeled as unitary noise); neither can be considered as probabilistic noise.
\\\medskip

Here, we examine how one might derive an accuracy threshold for surface codes exposed to general noise---one with no particular structure and includes probabilistic noise as a specific example---by generalizing existing analyses for probabilistic noise. There is perhaps a general expectation in the community that the same methods for probabilistic noise extend to general noise (this was true for older fault-tolerant schemes based on concatenated codes), and that the threshold numbers are similar. As we explain here, however, current known techniques for deriving an accuracy threshold for surface codes likely do not give a nontrivial (i.e., nonzero noise strength) threshold for general noise. While this is not a proof that fault-tolerant quantum computing based on surface codes cannot work under general noise, it serves as a caution that our current conclusions about surface codes, founded largely upon probabilistic noise studies, may not apply to actual quantum devices. This highlights the need for new ideas to resolve the question of whether a nonzero accuracy threshold exists for surface codes, and whether our confidence in the success of large-scale quantum computers built upon surface codes is well founded.

\section{Surface code basics}

\subsection{Code structure}

The (planar) surface code \cite{KitaevPlanar,Dennis,FreedmanMeyer} is an error-correcting code defined on a two-dimensional (2D) square lattice of qubits, some of which carry the computational data---the data qubits---and others are ancillary qubits---or just ancillas---used for the syndrome measurements in the error correction. The joint state of the data qubits, in the absence of errors, resides in a two-dimensional code space, thus encoding a single qubit of information, and is generally highly entangled across the different qubits. The encoded, or logical, qubit is thus delocalized over the entire 2D lattice.
\\
\medskip

\begin{figure}[!ht]
\centerline{\includegraphics[width=0.6\columnwidth]{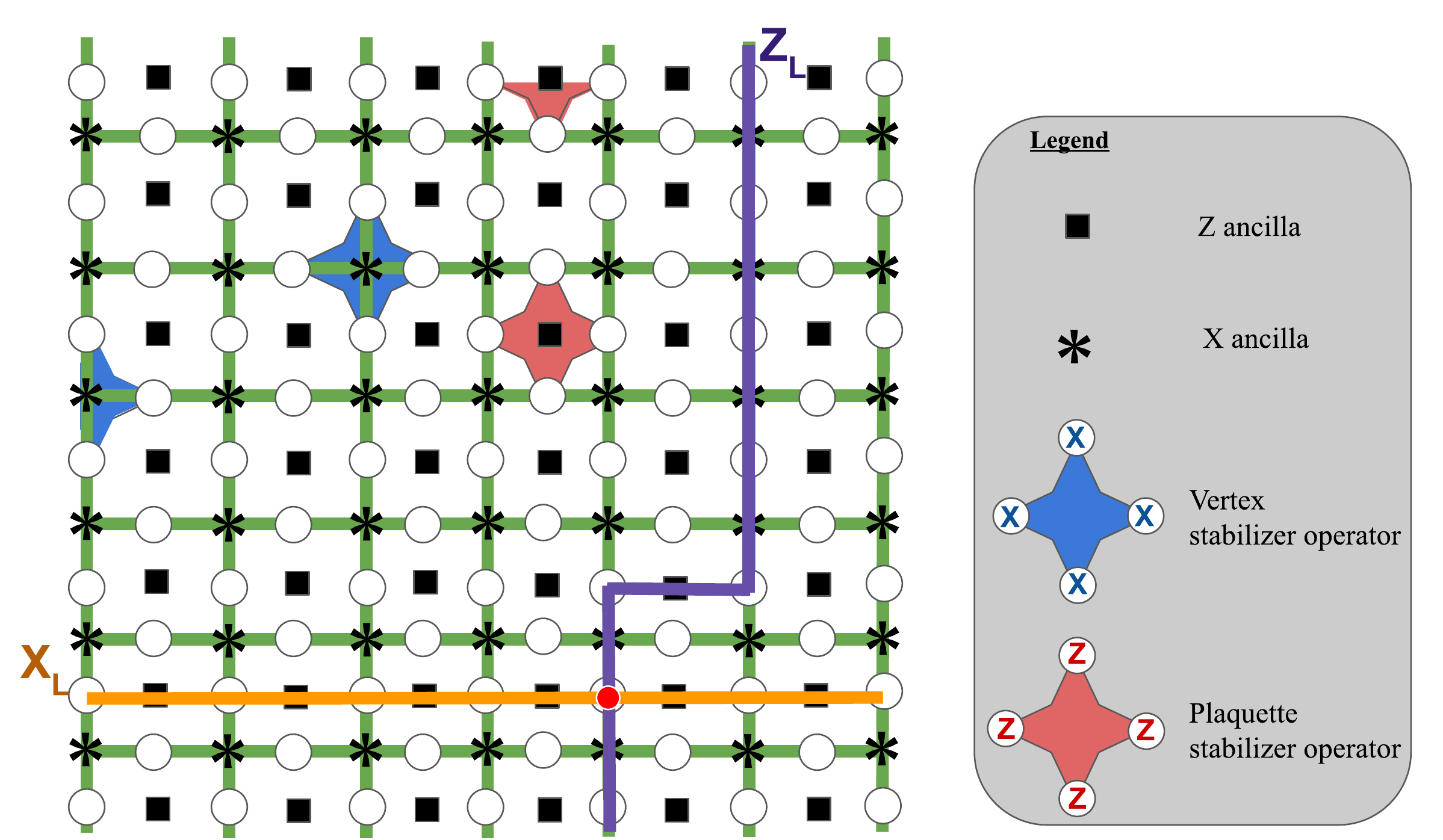}}
\caption{\label{fig:surface1}The \mbox{$L=7$} surface code lattice. Each set of qubits covered by a blue or red star indicates a measured stabilizer operator, namely, the projector \mbox{$\frac{1}{2}(\id \pm O_a O_b O_c O_d)$} for data qubits $(a,b,c,d)$ at the points of a star, and $O$ is $X$ or $Z$, depending on whether we are measuring an $X$- or $Z$-type stabilizer operator. Stabilizer operators are measured at the locations of all ancilla qubits. At the boundaries of the surface code, the stabilizer operators act on three (or two, at the corners) data qubits, instead of the usual four. A boundary made up of $X(Z)$ ancillas constitutes a ``smooth''(``rough'') boundary. The $X_\mathrm{L}(Z_L)$ operator consists of $X(Z)$ operators on the data qubits that forms a path that extends from one smooth(rough) boundary to another smooth(rough) boundary. The $X_\mathrm{L}$ and $Z_\mathrm{L}$ operators together generate the Pauli algebra of the encoded qubit. See, for example, Reference \cite{MartinisFowler} for further details.}
\end{figure}

The surface code lattice comprises, for odd $L$, $L-1$ rows and $L$ columns of vertices, such that there are two ``smooth" boundaries (left and right boundaries in Figure \ref{fig:surface1}), and two ``rough" boundaries (top and bottom boundaries in Figure \ref{fig:surface1}). There are altogether $L^2$ vertical edges and $(L-1)^2$ horizontal edges, and on each edge resides a data qubit, giving a total of $L^2+(L-1)^2$ data qubits. 
Ancillary qubits reside on vertices and in the centres of plaquettes, corresponding to two different ancilla types (see Figure \ref{fig:surface1}): $X$ ancillas that sit on the vertices, so named because they are for measuring the $X$-type stabilizer operators which act on the four (fewer, if at the lattice boundaries) data qubits surrounding the vertex; $Z$ ancillas at the center of the plaquettes for measuring the $Z$-type stabilizer operators that act on the four data qubits bordering the plaquette. There are $L(L-1)$ each of such $X$ and $Z$ ancillas, giving altogether $2L(L-1)$ ancillary qubits, or, equivalently, $2L(L-1)$ stabilizer operators. Altogether, there are $(2L-1)^2$ data and ancillary qubits in the lattice.\\
\medskip

The $2L(L-1)$ $X$- and $Z$-type stabilizer operators together specify a two-dimensional code space carried by the data qubits, namely, the state space on which the stabilizer operators act like the identity operation. The logical $X(Z)$ operator can be identified as, up to factors from the stabilizer group, the tensor product of $X(Z)$ on every data qubit in a set that connects one smooth(rough) boundary to the other (see Figure \ref{fig:surface1}). The code is designed to correct arbitrary errors on up to $t$ data qubits, with $t$ related to the lattice size $L$ as $t=\tfrac{1}{2}(L-1)$.\\
\medskip

\subsection{Recovering from errors}
The syndrome measurements, namely, the measurement of the $X$- and $Z$-type stabilizer operators, are carried out using the $X$ and $Z$ ancillas. The $X$-type stabilizers detect $Z$ errors on the data qubits, while the $Z$-type ones detect $X$ errors. Together, they detect arbitrary qubit errors, noting that $Y=XZ$ and that every qubit error can be written as a sum of $X$, $Y$, and $Z$. To measure one stabilizer operator, CNOT gates are applied consecutively, connecting each of the data qubits involved in that stabilizer operator to the same ancilla, effectively transferring the information about errors in those data qubits to the ancilla. The ancilla is then measured, and the measurement result, either $+1$ or $-1$, is recorded. A preliminary detection of an error happens when the ancilla measurement result differs from its value in the syndrome measurement  in the previous error correction cycle. Such a change in measurement value is referred to as a defect, is said to be located at the position of the ancilla, and is (in the ideal case) an indication that there are errors in the data qubits connected to that ancilla.\\
\medskip

\begin{figure}[!ht]
\centerline{\includegraphics[width=0.65\columnwidth]{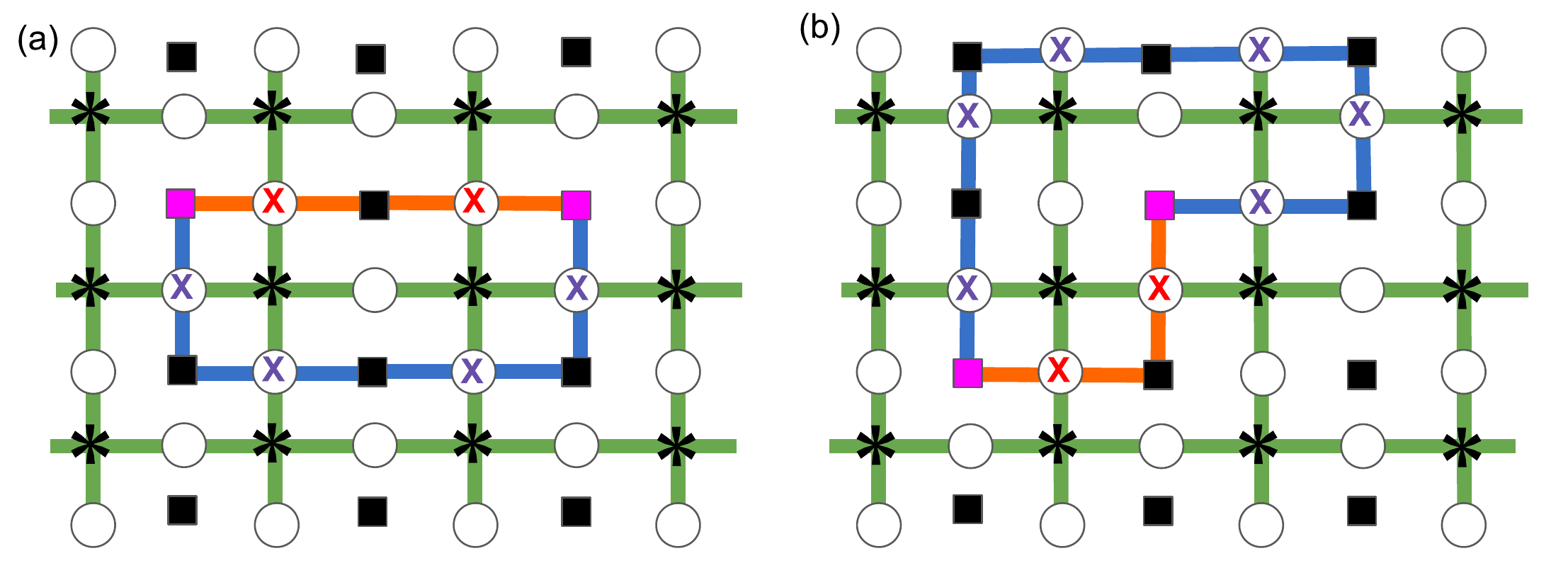}}
\caption{\label{fig:defects}Two examples of the link between errors and defects. Ancillas marked in magenta indicate the set of defects observed if the $X$ errors marked in red are the only errors. The same defect set is observed, if the errors are instead those qubits marked with purple $X$s. For a given set of defects, the decoder attempts to find a set of lattice paths that pair up the defects. A lattice path is a continuous line that joins adjacent $Z(X)$ ancillas. The blue and orange lines are possible lattice paths that pair up the observed defects. If the path passes through a data qubit, one identifies an $X(Z)$ error on that qubit.}
\end{figure}

The defect locations are gathered after a single round of syndrome measurements, i.e., measurement of all $2L(L-1)$ stabilizer operators, and the set of defect locations is processed, or decoded, to deduce what errors have occurred in the data qubits. The stabilizer operators are distributed across the lattice in a manner such that, if an $X$ error, say, occurred in a data qubit, the error flips the measurement results (e.g., a $+1$ to a $-1$) on the the two $Z$ ancillas neighboring it (one if it is on a lattice boundary). Both will manifest defects in the next round of syndrome measurement. 
If $X$ errors occur on two adjacent data qubits simultaneously, the $Z$ ancilla in between the two will encounter the flip twice, winding up with no defects, but the other two $Z$ ancillas next to the data qubits will register defects (see Figure \ref{fig:defects}). Analogous statements hold for $Z$ errors and associated $X$ ancillas. This gives the connection between defects and errors: Every error on a data qubit produces a pair of defects\footnote{The only exception to this is when an error occurs at the boundary of the code lattice. This will result in only one defect, as the other ancilla location is ``missing"---it falls outside of the boundary.}, and adjacent data qubits in error separate the defect pairs further apart. Defects manifest only on the ``outermost" ancillas, and the data qubits along the lattice path---a connected sequence of edges (see Figure \ref{fig:defects})---joining the two defects are in error. In fact, data qubits on any path with those same two defects as endpoints can be said to be in error, and this will be correct up to operators from the stabilizer group.\\
\medskip

This then suggests a way to decode the defects into errors. Connect pairs of defects together, and the data qubits on the connecting path are guessed to be in error. A path here refers to a connected line that joins two defects together; equivalently, we can regard a path as the set of edges that the connected line passes through (see Figure \ref{fig:defects}). This allows us to talk about data qubits that lie on a given path, and those data qubits are said to be in error, as deduced by the decoder according to the observed defects.\\
\medskip

Among the many possible connecting paths that join two defects, one chooses the path with the highest probability of error occurrence. If the errors occur independently, this translates into the shortest path, with the fewest data qubits in error. This is the basic principle of the minimum-weight perfect-matching (MWPM) decoder \cite{Dennis}, a standard decoder for the surface code, and the one we will use here \footnote{There are of course many other possible decoders for surface codes. In recent years, a variety of approaches to decoding the surface-code syndromes have been developed \cite{BravyiMLDecoder,MCMCdecoder,MultiPathDecoder,CellularDecoder,neural1,neural2,LinearTimeDecoder}, with varying levels of performance and advantages, e.g., ones that fail more often than say the MWPM decoder, but use less classical computation time, or ones that take inspiration from physical models in nature.}. See Figure \ref{fig:defects} for two illustrative examples---there, the lattice paths marked in orange are the shortest paths, and the ones that will be chosen by the MWPM decoder.\\
\medskip

The decoder chooses paths that connect observed defects in pairs. The recovery is then performed to (attempt to) reverse the errors that occurred. If the defects were for $X(Z)$ errors (i.e., observed on the $Z(X)$ ancillas), the $X(Z)$ operator is applied to all data qubits on the chosen paths. Now, if the actual errors that occurred, together with the applied recovery operations, form a connected path of data qubits that extends from one smooth(rough) boundary to the other, we wind up with an overall $X_\mathrm{L}(Z_\mathrm{L})$ operation on the encoded information, i.e., we have a logical error. The data qubit on every edge of such a path is acted upon by either by an error or the recovery operation, but not both (which would cause the path to become disconnected and no logical error results). We refer to such a path that connects the two relevant---i.e., smooth ones for $X$ errors, and rough ones for $Z$ errors---boundaries as a ``spanning path". Such spanning paths extend across the entire code lattice and correspond to logical errors.\\
\medskip

\subsection{Noisy recovery and the syndrome lattices}
In practice, the physical components used to implement the syndrome measurements---the CNOT gates, the ancillas, as well as the final measurements---will inevitably be noisy. This can lead to unreliable syndrome information, in the form of missing or spurious defects, that confuse the detection of errors. To obtain robust information in the presence of such imperfections, a standard procedure is to repeat the syndrome measurement a few times \cite{ShorFTQC}. Persistent erroneous syndrome information require multiple errors to have occurred over the multiple syndrome measurement cycles, which is unlikely. The total defect set gathered over the multiple cycles are then jointly decoded into a best guess for the actual error locations.\\
\medskip

Following past work on the subject (see, for example, Reference \cite{Dennis}), we assume the syndrome measurement is repeated $N\sim O(L)$ times before decoding is done. Often, $N=L$, but we do not need the specific form of $N$ here, only that it varies affinely with $L$. The defects gathered over the $N$ cycles can be visualized on a three-dimensional (3D) $L\times L\times N$ lattice, extending in two spatial ($L\times L$) and one temporal ($N$) directions. Each time slice corresponds to the surface code lattice (said to extend spatially) subjected to a single syndrome measurement cycle, and the defects occur at the ancilla locations where they manifest in space (where on the surface code lattice) and time (which syndrome cycle).
\medskip

This rectangular 3D lattice can be embellished into what we will refer to as a syndrome lattice (this was called ``a lattice of dots and lines" in Reference \cite{FowlerProof}), useful for the MWPM decoder and for our discussion below. We first consider the situation of $Z$ errors occurring only\footnote{The surface code deals with $X$ and $Z$ errors separately, so we can treat them separately. Any arbitrary error can be decomposed into a sum of $X$, $Y=XZ$, and $Z$ errors, so this handles all errors.}. The vertices of the 3D lattice described above are the locations of the $X$ ancillas, and hence locations where defects can manifest when $Z$ errors occur. Two adjacent vertices on the same time slice are connected by an edge, on which a data qubit resides. If a single fault (i.e., something going wrong) occurs on the data qubit, causing a $Z$ error, the two ancillas on the two vertices will manifest defects. \\
\medskip

We want to extend this feature that an edge joins two vertices if a single fault, nominally said to occur on that edge, can lead to defects manifesting on the vertices to faults that arise in the course of error correction. The ancillary qubits themselves can also have faults, and a $Z$ error on an $X$ ancilla will cause two defects to appear in two adjacent time slices (first time slice when the error occurs, and the next, when the ancilla is reset and hence the error is eliminated) on the same ancilla location. We add an edge that joins the two vertices---same spatial location of the ancilla but separated one time step apart---where the two defects manifest, and that (temporal) edge is said to have a fault when that $X$ ancilla suffers a $Z$ error.\\
\medskip

\begin{figure}
\centerline{\includegraphics[width=0.8\columnwidth]{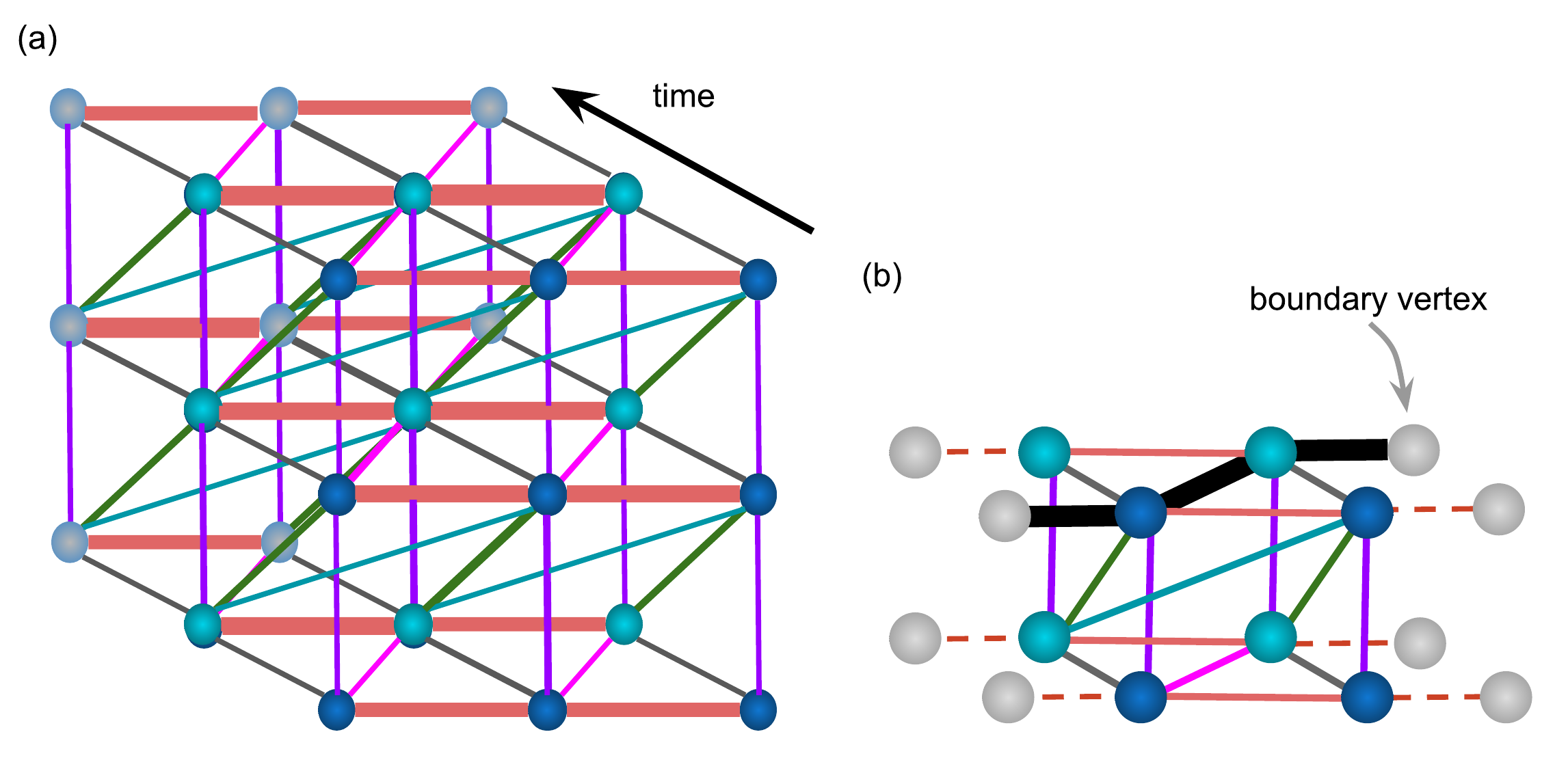}}
\caption{\label{fig:unitLattice}(a) A depiction of eight unit cells of the syndrome lattice for $X$ errors, with the vertices representing the $Z$ ancillas. Each edge connects a possible defect pair that can be observed from a single fault, and the colors of the edges indicate the type of fault that can occur. For example, each gray edge connects the defects that result from a faulty measurement---the pair of defects occur at the same spatial location but at consecutive times. (b) A small syndrome lattice for $X$ errors for a $2\times 3$ surface code, with two syndrome measurement cycles ($N=2$). The accompanying relevant boundary vertices are shown, as well as a possible spanning path, illustrated here as the set of thick black edges.}
\end{figure}

Similarly, when a CNOT gate in the syndrome measurement circuit has a fault, it causes $Z$ errors on the ancillary qubit and/or data qubit involved, and possibly spread to other qubits by subsequent gates (if any). Defects can then manifest at a pair of vertices, and we can again join them together by an edge. A CNOT fault is said to occur on that edge. A detailed study \cite{WangFowler} (see also Reference \cite{Autotune} for how this information can be incorporated into the decoder) of how faults can lead to defects allows us to put an edge between two vertices of the 3D lattice, wherever a single fault can cause a pair of defects to manifest at the vertices joined by the edge. The resulting lattice, now with many more edges than the original 3D lattice, is called the syndrome lattice for $Z$ errors; similarly, one can construct a syndrome lattice for $X$ errors, identical to that for $Z$ errors, but now with $Z$ ancillas at the vertices. Figure \ref{fig:unitLattice} shows a portion of a syndrome lattice, with each interior vertex connected by edges to 12 other vertices. 

\medskip

With this picture of syndrome lattices, the decoding of defects into errors can be carried out in a manner similar to the no-noise case, except that the paths connecting defects now traverse the 3D lattice, possibly extending across different time slices. Edges on the paths are now edges on the syndrome lattice, not those on the original surface code lattice. A spanning path in the syndrome lattice for $X(Z)$ errors is now one that starts on one relevant boundary \emph{face} (comprising all the relevant boundaries of the different time slices) of the syndrome lattice and ends on the opposite face. A logical error still results when the errors and recovery form such a spanning path \footnote{In principle, one can also talk about paths that start and/or end on the temporal boundaries, i.e., the first and last time slices of the syndrome lattice. For some decoding protocols, such paths can lead to logical errors in the next round of syndrome measurement cycles. We ignore such complications here, as they do not change our argument about the accuracy threshold.}.

\subsection{Error description}\label{sec:errorModel}
We also need a description of the noise that affects our quantum computer. Without loss of generality, noise in the quantum computer, over some discrete time interval (e.g., computational clock cycle) can be described by completely positive (CP) and trace-preserving (TP) maps acting on the data and ancillary qubits, as well as any environmental degrees of freedom that come into play. This can arise from background noise, or be the result of noisy gates acting on subsets of qubits. Noise arising from an imperfect gate is assumed to affect only those qubits participating in that gate.\\
\medskip

The noise description can be easily incorporated into our discussion using the syndrome lattices: Each edge of the syndrome lattices is acted upon by a CPTP map $\cN_a$, for $a=1,2,\ldots 2A$ numbering the edges in the two syndrome lattices, and $A$ is the total number of edges in a single lattice. Each $\cN_a$ acting on an edge in the $X(Z)$ lattice is assumed to be such that it can lead to $X(Z)$ error(s).
$\cN_a$ can be a primitive noise process like background noise on a single ancilla (such an $\cN_a$ acts on a temporal edge in the syndrome lattices), but also propagated noise like a fault occurring on a CNOT acting on a pair of qubits and the resulting errors are spread by subsequent gates to other qubits before the syndrome measurement is done. Since the primitive noise is CPTP, so will be the propagated noise, and the propagated noise, by construction, is associated with an edge in the syndrome lattices. \\\medskip

We split each noise map $\cN_a$ into two parts in the following manner \cite{Ben-Or}. Consider the deviation of $\cN_a$ from the identity map $\id$, and let $\eta_a\equiv\Vert\cN_a-\id\Vert$. Here, we use a unitarily invariant and submultiplicative superoperator norm such that $\Vert\cE\Vert=1$ for any CPTP $\cE$ (e.g., the diamond norm works). We define $\eta\equiv\max_a\eta_a$. Then, $\Vert\cN_a-\id\Vert\leq\eta$ $\forall a$, and we can write each $\cN_a$ as
\begin{equation}\label{eq:Nk}
\cN_a=(1-\eta)\id+\cF_a.
\end{equation}
with $\Vert \cF_a\Vert\leq 2\eta$ for all $a$, and $\eta$ characterizes the strength of the general noise. We refer to $\cF_a$ as a ``fault", i.e., something goes wrong that can cause errors on the qubits associated with that edge, and can lead to defects at the endpoints of the $a$th edge. The $\id$ term (with weight $1-\eta$) is regarded as the ``no-fault" situation, and does not cause defects to appear.\\
\medskip

For probabilistic noise, $\cF_a$ can be written as $\eta\cE_a$ where $\cE_a$ is a CPTP map, so that $\cN_a$ carries the interpretation that nothing happens (i.e., $\id$ is applied) with probability $1-\eta$ and a fault $\cE_a$ happens with probability $\eta$. For general noise, $\cF_a$ need not be proportional to a CPTP map, and there is no such straightforward probabilistic interpretation.

\section{Accuracy threshold for probabilistic noise}
Here, we derive the accuracy threshold for the case of probabilistic noise, starting with the original basic argument of Reference \cite{FowlerProof} (which provided the main existing analytical proof for surface codes under probabilistic noise), and extending it to include higher-order corrections.

\subsection{Basic argument}
Let us first review the logic leading to the accuracy threshold estimate for surface codes as originally presented in Reference \cite{FowlerProof}. We assume a probabilistic noise model as described in Section \ref{sec:errorModel}: Each edge of the syndrome lattice has probability $p$ of having an error that can result in defects at its endpoints. $p$ here is the $\eta$ of Equation \eqref{eq:Nk}, but we write $p$ to emphasize its probabilistic interpretation. To deduce the threshold for surface codes, we want to bound the total probability that things can go wrong, i.e., that logical errors occur after the recovery operation as determined by the MWPM decoder. As described earlier, a logical error occurs when the errors that occur, together with the recovery operations, form a spanning path in the syndrome lattice. Such a path has a number of edges---its length $r$---no fewer than $L$ in order to connect two opposing boundaries. At least $\lceil r/2\rceil$ of those $r$ edges are associated with errors, or the pairing of the defects on the path, as decided by the MWPM decoder, cannot be minimum weight. The probability that the errors and resulting recovery operations form a specific length-$r$ path is
\begin{equation}\label{eq:Pr}
P(r)\equiv \sum_{k=\lceil r/2\rceil}^r\binom{r}{k}p^k(1-p)^{r-k}\leq p^{\lceil r/2\rceil}\sum_{k=0}^r\binom{r}{k}=p^{\lceil r/2\rceil}2^r,
\end{equation}
where we assume the $k\in\bigl[\lceil r/2\rceil,r\bigr]$ erroneous edges can occur anywhere on the path (each with probability $p$), and the remaining edges have no errors (each with probability $1-p$, and added to the path by the recovery).\\\medskip

We want to estimate the total probability of occurrence of such spanning paths, giving the total probability of logical errors. It is difficult to count all such paths exactly. However, one can arrive at an upper bound in the following way, as originally proposed in Reference \cite{FowlerProof}. We begin by choosing a vertex on one of the relevant boundary faces of the syndrome lattice---there are $2NL$ \footnote{This $2NL$ can be understood as follows: The spanning path can extend from one relevant boundary face to the opposite face. The starting vertex can hence be one of $NL$ vertices, multiplied by 2 for the two syndrome lattices.} such possible starting vertices. From this starting point, we choose the next vertex---there are no more than 11 possible vertices to choose from, given the structure of the syndrome lattice. We continue adding vertices to the path, each time choosing a next vertex connected by an edge to the previous one---there are 11 options each time, excluding the vertex we just came from---to arrive at a path with length $r$: $r+1$ vertices in total, connected by $r$ edges. We consider all such length-$r$ paths, which includes spanning paths that correspond to logical errors, but also many paths that start at one boundary face but never reach the opposite boundary face. With this, we can upper-bound the total probability of logical errors by
\begin{equation}\label{eq:probthreshold}
\sum_{r=L}^\infty(\textrm{number of such length-$r$ paths})P(r)\leq cNL \frac{(22^2p)^{\lceil L/2\rceil}}{1-22^2p}\equiv \PL,
\end{equation}
where $c$ is the numerical constant $c\equiv 2 (23/22)\simeq 2.1$. The upper bound expression $\PL$ expression can be arrived at by straightforward counting, together with Equation \eqref{eq:Pr} (see the Methods section). Reference \cite{FowlerProof} concludes by saying the $\PL$ can be made arbitrarily small by increasing $L$, as long as $p < 1/(22^2)=1/484\equiv p_\mathrm{th}$. This ensures $P_\mathrm{UB}(L+1)\leq P_{UB}(L)$, and $p_\mathrm{th}$ can be identified as the accuracy threshold for the surface code in the presence of probabilistic noise.\\
\medskip 

There are some caveats, however, to this analysis. Observe that $\PL$ is unphysical --- it is larger than 1 --- as a probability for $p$ below, but close to, the threshold value.
That this upper bound $\PL$ can be larger than 1, when the actual logical probability cannot be so, is simply a consequence of the overcounting in the argument. There are two sources of overcounting: (1) In deriving Equation \eqref{eq:Pr}, $P(r)$ is itself an over-count since not all possible allocations of $k\leq \lceil r/2\rceil$ erroneous edges in a path of length $r$ lead to a logical error after the recovery. In fact, the surface code correctly removes the errors in many situations with larger than $t$ errors. Furthermore, our bound on $P(r)$ itself contains an inequality. (2) We were unable to exclude from our counting paths that started on one boundary face but did not reach the opposite boundary face. (2) no doubt gives a much larger contribution to the overcounting than (1).\\
\medskip

Also, note that we are only able to say that, when $p<p_\mathrm{th}$, the \emph{upper bound} $\PL$ can be made smaller and smaller as $L$ grows. This does not technically mean that the actual logical error probability shrinks as $L$ grows, just that our upper bound on it shrinks. It does, nevertheless, suffice to guarantee that, for a given $p$ below the threshold value, we can find a large enough $L$ such that the logical error probability, which is smaller than $\PL$, is small enough. This statement is true also of the standard accuracy threshold estimates for concatenated codes (see, for example, References \cite{Ben-Or,AGP,TerhalBurkard}), though in those cases, the upper bounds are likely quite close to the actual logical error probability, unlike the surface-code situation here.\\
\medskip

\subsection{Higher-order terms: the disconnected pieces}
There is, however, a more worrying issue with the analysis of Reference \cite{FowlerProof}. On the one hand, we mentioned that the analysis overcounts by including many paths that do not span the lattice and hence do not contribute to the logical error probability; on the other hand, we now point out that it in fact misses many paths that do contribute.\\
\medskip

\begin{figure}[!ht]
\centerline{\includegraphics[width=0.8\columnwidth]{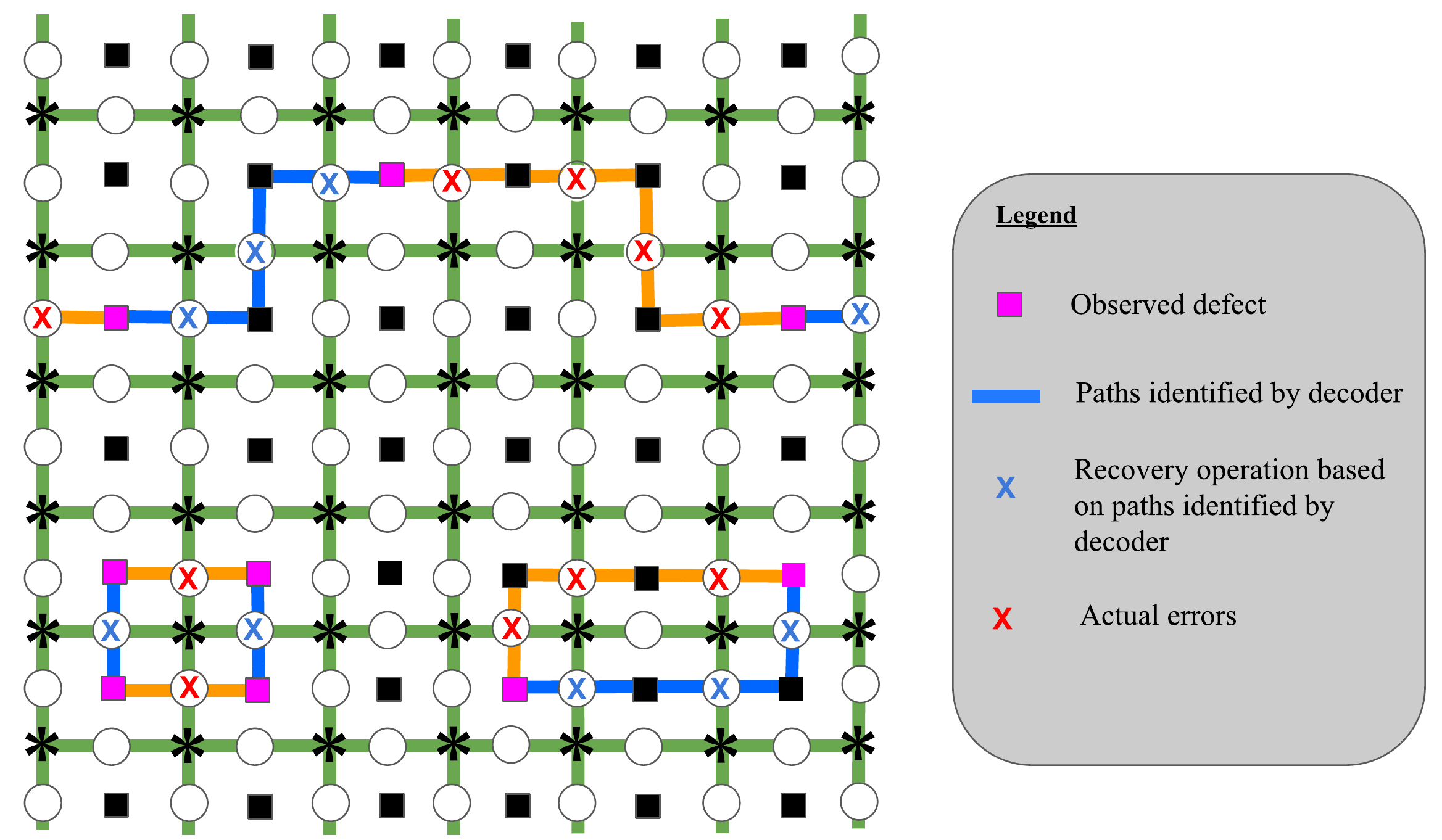}}
\caption{\label{fig:loopspath}An example where errors on data qubits give rise to defects as observed on the surface code lattice. The decoder determines the paths that pair up observed defects, and recovery operations are applied to the data qubits on those paths. In this case, a spanning path made up of errors and erroneous recovery operations occurs, ultimately causing a logical $X_L$ error. Note that in this example, there are a total of 10 edges associated with errors and 9 edges associated with recovery operations. Each edge associated with error comes from a single fault \footnote{More precisely, multiple single-fault locations could give the same defect pair pattern and thus identify with the same edge. Hence, each edge that is associated with error could in principle come from an \emph{odd} number of faults. However, this is only a minor detail that does not change the rest of our conclusions.}, so this example occurs with a probability of $p^{10}$. However, the spanning path alone only has 5 edges associated with errors, which by itself occurs with probability $p^5$. Simply counting spanning paths is insufficient as one needs to consider all higher-order loops.}
\end{figure}

The argument of Reference \cite{FowlerProof} counts all connected paths that span the lattice and can lead to a logical error. One could also have disconnected paths alongside a connected one (see an example in Figure \ref{fig:loopspath}). These disconnected paths may lead to no additional logical errors---they can correspond to errors that occur but are then correctly removed by the recovery procedure, or they may also span the lattice---but give corrections to the probability of occurrence of the basic connected path. Each disconnected path contributes a higher-order term, but there are many more possibilities of such disconnected pieces, and the combinatorial factor could potentially offset the higher-order probability factor. None of these disconnected pieces were considered in the analysis of Reference \cite{FowlerProof}.\\
\medskip

\medskip

It turns out, however, that these disconnected paths do not change the final bound for the situation of probabilistic noise, as we now explain. We simply need to understand the correction factor, associated with each length-$r$ spanning path, needed to include the disconnected paths. This can be counted as follows. A particular length-$r$ connected path, chosen as described in the previous subsection, corresponds to a particular set of $r$ edges, out of the $A$ edges in the syndrome lattice. Disconnected paths can be included alongside these chosen $r$ edges by deciding if the remaining $A-r$ edges have errors or not. If we count all possibilities, the correction factor, to be multiplied to $P(r)$ of Equation \eqref{eq:Pr}, is then
\begin{equation}
\sum_{\ell=0}^{A-r}\binom{A-r}{\ell}p^\ell(1-p)^{A-r-\ell}=[p+(1-p)]^{A-r}=1.
\end{equation}
Here, the summation index $\ell$ is the number of edges not on the spanning path ($A-r$ of these) that are chosen to have errors; the remaining $A-r-\ell$ edges are error-free.
That the higher-order disconnected correction amounts to a multiplicative factor of 1 explains why the analysis of Reference \cite{FowlerProof} still gives the right answer, despite disregarding these possibilities.\\
\medskip

This same argument, however, does not work when one attempts to extend the analysis to general noise, as we will see in the next section.

\section{Accuracy threshold for general noise}

To see where the problem lies for general noise, let us organize the analysis in a more transparent manner. Let us regard the entire surface code error correction procedure---the multiple rounds of syndrome extraction, the decoding, and the recovery---as a single quantum computation $\cir$, built from noisy operations. $\cir$ can be written as a sequence of CPTP maps,
\begin{equation}
\cir\equiv \sR\circ\sM\circ\sG.
\end{equation}
Here, $\sG$ refers to all the gate operations (the CNOT gates) for the $N$ rounds of syndrome extraction, as well as any memory (identity gate) steps, $\sM$ collects all the measurements on the ancilla qubits to extract the syndromes, and $\sR$ is the decoding procedure followed by the recovery operations determined by the decoding. $\sM$ refers only to the measurement part of the syndrome extraction procedure, i.e., it does not include the CNOT gates that transfer the error information from the data qubits to the ancilla qubits---those are collected in $\sG$. All measurements from the multiple syndrome extraction rounds are assumed to be delayed to the end, imitating the actual situation of measuring as we go along by assuming no fault insertions between the last CNOT on the ancilla qubit (in $\sG$) and the actual measurement (in $\sM$). The noise in the measurement is assumed to be lumped together with the noise of that last CNOT gate. $\sM$ can thus be considered as perfect, i.e., no noise. $\sR$, which describes the procedure of deducing the errors from the defect information collected over the multiple syndrome rounds, followed by the standard ``virtual" recovery using a Pauli-frame change, is actually a purely classical procedure, and hence can also be considered noise-free.\\
\medskip

All the noisy operations are then only in $\sG$, which itself can be written as a sequence of noisy gates. Now, the gate sequence in $\sG$ comprises only error correction gates---we are not doing any nontrivial computation. In fact, recalling how the syndrome lattices are constructed, which takes the sequence of error correction gates already into account, $\sG$ can be written as a sequence of the CPTP maps $\cN_a$ [recall Equation \eqref{eq:Nk}], one on each edge of the two syndrome lattices. Thus, we have,
\begin{equation}\label{eq:G}
\sG=\cN_{2A}\circ\cN_{2A-1}\circ\cdots\circ\cN_1.
\end{equation}
Inserting the fault/no-fault split of $\cN_a$ into $\sG$, we can then read $\sG$ as a sum over ``fault paths", each comprising a sequence of $\id$ and $\cF$s (i.e., the $\cF_a$s, dropping the index for brevity),
\begin{equation}\label{eq:FP}
\sG=\sum\textrm{fault paths}=(1-\eta)^{2A}\id\circ\cdots\circ \id+(1-\eta)^{2A-1}\id\circ\cdots\circ\id\circ\cF_1+\ldots\
\end{equation}
Each fault path corresponds to the insertion of $\id$---carrying weight $(1-\eta)$---or fault $\cF$ on each edge of the syndrome lattice. The very first term of Equation \eqref{eq:FP} is proportional to the identity map, as it should be for error correction operations when no faults occur.\\
\medskip

We can split $\sG$ into two pieces, one ``good", the other ``bad": 
\begin{equation}
\sG\equiv \sG_\mathrm{good}+\sG_\mathrm{bad}.
\end{equation}
The good piece $\sG_\mathrm{good}$ comprises the sum of all the fault paths that, after passing through $\sM$ and $\sR$, result in no errors, i.e., the errors that occur are correctly removed by the error correction \footnote{Actually, the good piece can contain also fault paths where the errors are not corrected in this decoding round but fixed only in the next round. Such situations occur in decoders that allow errors to be ``joined to the temporal boundary" and delayed to be fixed in the next round. Just as we have been ignoring spanning paths that start or end on temporal boundaries, we will ignore these here. Again, they do not change our arguments.}; the bad piece $\sG_\mathrm{bad}$ is the sum of those fault paths---the remaining ones---that result in a logical error after $\sM$ and $\sR$. What we want to do is to obtain an upper bound on the size of $\sG_\mathrm{bad}$, as a bound on how badly the error correction can fail.\\
\medskip

\subsection{An upper bound}
How do we identify which fault paths in $\sG$ belong to $\sG_\mathrm{bad}$? These are precisely those that, after $\sM$ and $\sR$, have at least one spanning path, comprising edges associated errors or recovery operations. Reference \cite{FowlerProof} gave a way of counting (in fact, \emph{over}counting) such paths, as explained in the previous section---that counting does not care whether we have probabilistic or general noise. 
For each such identified path $\cP$, with specified error and recovery edges, there is a corresponding error-only fault path in $\sG_\mathrm{bad}$, with the fault $\cF$ insertions precisely on those edges in $\cP$ with errors, and identity on all the other edges in $\cP$ (these become the recovery edges in $\cP$ only after $\sM$ and $\sR$). What about the other edges not in $\cP$? Those can, in principle, have fault insertions or not and can be accounted for by inserting the full $\cN$ (again, dropping the index $a$ for brevity), i.e., both possibilities, $\id$ and $\cF$, hence taking care of the ``disconnected pieces" mentioned earlier. We write the sum of all such paths symbolically as
\begin{equation}\label{eq:GB}
\sG'=\sum_\cP \cF(\cP)\circ\cN(\overline{\cP}),
\end{equation}
where $\cF(\cP)$ denotes inserting $\cF$ or $\id$ on the edges in $\cP$ as described above, and $\cN(\overline\cP)$ denotes inserting $\cN$ on every edge of the syndrome lattice not in $\cP$, i.e., the complement of $\cP$, written here as $\overline{\cP}$.\\
\medskip

Note that we have written $\sG'$ in Equation \eqref{eq:GB}, not $\sG_\mathrm{bad}$, for the following reason: $\sG'$ contains all fault paths that translate into logical errors, i.e., contains $\sG_\mathrm{bad}$, but also ones that do not ultimately give logical errors. Part of this comes from the overcounting of $\cP$---as mentioned earlier, the counting in Reference \cite{FowlerProof} includes many $\cP$s that are not spanning paths. In addition, our way of adding fault insertions in $\overline\cP$ by inserting the full $\cN$ could give arrangements of errors such that, after passing through $\sM$ and $\sR$, lead to a recovery that do not realise $\cP$ as the error-plus-recovery path, but rather connect the resulting defects in a manner that do not give a spanning path and hence no logical error (see the Methods section for an example and further elaboration on this point). This means that we have included some---possibly very many---fault paths originally supposed to be in $\sG_\mathrm{good}$ within $\sG'$.\\
\medskip

Despite this, let us persist with $\sG'$ for a bit more, before coming back to discuss the consequences of the overcounting. The goal is to bound the part of $\cir$ that leads to logical errors, i.e., the bad piece, $\sR\circ\sM\circ\sG_\mathrm{bad}$. We do not yet have $\sG_\mathrm{bad}$; instead, let us first compute the norm of $\sR\circ\sM\circ\sG'$. Using a submultiplicative superoperator norm such that any CPTP map has unit norm, we have 
\begin{equation}\label{eq:sumFP}
\Vert\sR\circ\sM\circ\sG'\Vert\leq \Vert\sR\Vert\,\Vert\sM\Vert\,\Vert\sG'\Vert=\Vert\sG'\Vert\leq \sum_\cP\Vert\cF(\cP)\Vert\,\Vert\cN(\overline\cP)\Vert=\sum_\cP\Vert\cF(\cP)\Vert.
\end{equation}
This final piece $\sum_\cP\Vert\cF(\cP)\Vert$ can be bounded in a similar way as in the probabilistic noise case discussed above, replacing $p$ with $2\eta$ ($\geq \Vert\cF_a\Vert\forall a$), and we find (see Methods section),
\begin{align}\label{eq:Wprime}
\Vert\sR\circ\sM\circ\sG'\Vert\leq \sum_\cP\Vert\cF(\cP)\Vert\leq cNL\frac{(22^2\cdot 2\eta)^{\lceil L/2\rceil}}{1-22^2\cdot 2\eta}\equiv W'_{\mathrm{UB}}(L).
\end{align}
For probabilistic noise, $\WpL=\PL$, our upper bound from before, if we read $2\eta$ as $p$. For general noise, we can draw the same conclusion as before, that $\WpL$ shrinks as $L$ increases if $2\eta\leq 1/22^2$.

\subsection{Bounding the bad piece}
So, what does this upper bound, established using $\sG'$, tell us about the actual quantity we care about, namely, the occurrence of logical errors, which involves $\sG_\mathrm{bad}$ instead? For probabilistic noise, this is straightforward. The sum over fault paths in $\sG'$ or $\sG_\mathrm{bad}$ is a probabilistic sum, with each additional term contributing a positive quantity. The overcounting in $\sG'$ compared to $\sG_\mathrm{bad}$ hence only makes its norm larger. We can hence say that, for probabilistic noise,
\begin{equation}
\Vert\sR\circ\sM\circ\sG_\mathrm{bad}\Vert\leq\Vert\sG_\mathrm{bad}\Vert\leq \Vert\sG'\Vert\leq \WpL,
\end{equation}
recovering the conclusions for probabilistic noise discussed earlier [with $2\eta$ in place of $p$, so that $\WpL=\PL$]. As before, we see that the fault insertions on $\overline \cP$ contribute trivially: They became 1 after taking the norm of $\cN(\overline\cP)$ in $\sG'$.\\
\medskip

For general noise, however, we run into difficulties. Each term in a sum of fault paths can no longer be said to always give a positive contribution. Plus and minus signs, or even generally complex phases, associated with each term can lead to cancellations and destructive interference. We cannot then claim that $\sG'$, which contains fault paths that do not lead to logical errors, has a larger norm than $\sG_\mathrm{bad}$ itself. In fact, that the so-called disconnected pieces, encapsulated in the $\cN(\overline\cP)$ term in $\sG'$, disappear from the final bound on $\sG'$ only because the $\id$ and $\cF$ terms on every edge are there to ensure that the full CPTP $\cN$ appears in the sum and hence has unit norm. However, not all of the terms will lead to logical errors, so in $\sG_\mathrm{bad}$ itself, we do not expect the full $\cN$ to always occur (see an elaboration on this point in the Methods section), and these disconnected pieces should appear nontrivially in $\sG_\mathrm{bad}$, with incomplete cancellations.\\
\medskip

One can take the opposite tack and bound each fault path separately before taking the sum, so that no such cancellations can appear and each additional fault path contributes positively, as in the probabilistic noise situation. If we do this, that we are overcounting no longer matters. Then,
\begin{equation}\label{eq:sumFP2}
\Vert\sR\circ\sM\circ\sG_\mathrm{bad}\Vert\leq \sum_\cP\Vert\cF(\cP)\Vert\sum_{a\in\overline\cP}{\left[(1-\eta)\Vert\id\Vert+\Vert\cF_a\Vert\right]},
\end{equation}
where we have taken the norm of the individual $\id$ and $\cF$ terms in $\cN$ before taking the sum, to avoid  any possible cancellations. Noting that $\Vert\cF_a\Vert\leq 2\eta$ and following our earlier argument (see Methods section), we have
\begin{align}\label{eq:W}
\Vert\sR\circ\sM\circ\sG_\mathrm{bad}\Vert&\leq cNL(1+\eta)^A\frac{{\left[\frac{22^2\cdot2\eta}{(1+\eta)^2}\right]}^{\lceil L/2\rceil}}{1-\frac{22^2\cdot 2\eta}{(1+\eta)^2}}\equiv \WL\,,
\end{align}
with $c=2(23/22)$ as before.\\
\medskip

The real situation lies somewhere between these two extremes of $\WL$ and $\WpL$. Some of the disconnected terms probably do appear together so that the full $\cN$ appears in $\sG_\mathrm{bad}$, but this probably does not occur in all cases. How to distinguish those cases remains a difficult counting problem at the moment.\\
\medskip

Our analysis here with $\sG'$ actually mirrors similar considerations in the old proofs of fault-tolerant quantum computing with concatenated codes and recursive simulation (see, for example, Refs.~\cite{Ben-Or, AGP}. There, however, their corresponding $\sG'$ term is actually just $\sG_\mathrm{bad}$, as every term identified genuinely gives a logical error. No such mixing of terms over from the good $\sG_\mathrm{good}$ side happens there. Again, our difficulty here lies in not being able to precisely count only the paths that lead to logical errors. That first counting lattice-spanning paths and then adding disconnected pieces does not suffice here is a direct consequence of the fact that the surface code decoding requires a \emph{global} consideration of all defects that appear. One simply cannot decide whether a logical error results just from looking at a subset of defects or erroneous edges.

\subsection{A cost-benefit analysis}
As observed earlier, the upper bound $\WpL$ shrinks as $L$ increases, as long as $\eta$ is small enough, establishing a threshold condition on $\eta$. For $\WL$, there is no such threshold on $\eta$: $A$ grows as $L^3$, and the $(1+\eta)^A$ factor grows more rapidly than the $\eta^{\lceil L/2\rceil}$ factor shrinks as $L$ increases, so that $\WL$ blows up unless $\eta=0$. Thus, we cannot establish a nontrivial (i.e., $\eta>0$) accuracy threshold condition using the $W_\mathrm{UB}$ upper bound on $\sG_\mathrm{bad}$, while $\WpL$ is not actually an upper bound on $\sG_\mathrm{bad}$. \\
\medskip

This is, of course, not a proof that there is no nontrivial accuracy threshold for surface codes in the presence of general noise. A better proof method could derive an upper bound on $\sG_\mathrm{bad}$ that does yield a nontrivial threshold. In our current approach, which generalizes existing proofs to include the disconnected pieces, the disconnected pieces seem to be the ones causing the problem---our simplistic counting in $\WL$ led to the $(1+\eta)^A>1$ factor that explodes as $L$ grows. However, as we now attempt to argue, just by improving the counting of these disconnected pieces alone will likely not solve the problem.\\
\medskip

Observe that both $\WL$ and $\WpL$ take the form,
\begin{equation}\label{eq:genExp}
\underbrace{f(\alpha,\eta)(1+\alpha\eta)^{v(L)}}_{C(L)}\underbrace{{\left[\frac{22^2\cdot2\eta}{(1+\alpha\eta)^2}\right]}^{\ell(L)}}_{B(L)},
\end{equation}
where $\ell(L)\equiv\lceil L/2\rceil$, a function growing linearly with $L$; $v(L)\equiv A\sim L^3$, a function growing with the volume of the syndrome lattice; $f(\alpha,\eta)\equiv cNL{\left[1-\frac{22^2\cdot2\eta}{(1+\alpha\eta)^2}\right]}^{-1}$; and $\alpha$ is a parameter such that $\alpha =0$ for $\WpL$ while $\alpha=1$ for $\WL$. Expression \eqref{eq:genExp} can be split into two factors, $C(L)$ and $B(L)$, as indicated above (suppressing the $\alpha$ and $\eta$ dependences for brevity). $B(L)$ can be identified as the benefit of doing error correction with surface codes---as $L$ increases, the code is capable of removing a number of errors that grows linearly with $L$, so that the remnant ``bad" piece shrinks as $\sim\eta^{\ell(L)}$, as opposed to just $\eta$ without error correction. $C(L)$ represents the cost of doing error correction, collecting together the terms that capture the fact that there are more locations for faults to occur as $L$ grows. Written this way, the presence of an accuracy threshold can be thought of as arising from a cost-benefit analysis. A nontrivial threshold exists when the benefit outgrows [i.e., $B(L)$ shrinks] the cost as the scale $L$ of the code increases. For $\WL$, the cost $C(L)$ grows only linearly with $L$---it has no dependence on $v(L)$---while $B(L)$ shrinks exponentially with $L$, giving a nontrivial threshold; for $\WpL$, however, $C(L)$, with the exponential $L$ dependence on $v(L)$, quickly outstrips the exponential suppression in $B(L)$, and no nontrivial threshold exists.\\
\medskip

As argued earlier, the actual norm of $\sG_\mathrm{bad}$ should lie somewhere in between $\WL$ and $\WpL$. If one is able to improve the counting of the disconnected pieces, by identifying when the $\id$ and $\cF$ terms of $\cN$ occur together or not, the same logic as we have followed in this work will give a bound on $\sG_\mathrm{bad}$ that has a similar form as Equation \eqref{eq:genExp}, but with an $\alpha$ value somewhere between 0 and 1. Whatever value $\alpha$ turns out to be, however, as long as it is nonzero, the dependence on $v(L)\sim L^3$ will appear in the bound, causing the cost $C(L)$ to grow much more rapidly then the suppression provided by the benefit term $B(L)$ which shrinks only with an exponent $\ell(L)\sim L$, yielding again no nontrivial threshold.\\
\medskip

It thus appears difficult to escape this conclusion of a trivial threshold condition following our current lines of proof. It may suggest that there is no nontrivial threshold for general noise, beyond the earlier simplistic probabilistic noise considerations, but at least, our argument here suggests that a genuinely new idea---different from overcounting first spanning paths and then adding in disconnected pieces, reminiscent of past work on such threshold analyses---is needed to have the hope of deriving a nontrivial threshold. It may very well be that one has to be able to identify only fault paths that genuinely lead to logical errors, i.e., only those in $\sG_\mathrm{bad}$, but this appears to be a very challenging task.

\section{Conclusions}
The surface code has emerged as a strong contender for building large-scale quantum computing, with its promise of experimental requirements that are more feasible, including the general consensus of a less stringent fault-tolerance threshold. That consensus, however, is largely founded upon studies that modelled noise in a probabilistic manner, but unfortunately do not encompass all noise seen in actual quantum devices. The expectation is that the threshold conclusions can be extended, using similar proof techniques, to include general noise, as was the case for older fault-tolerance schemes based on concatenated codes like the Steane code.\\
\medskip

Here, our attempt to do precisely that, namely, to extend existing arguments that gave the surface code threshold under probabilistic noise to the case of general noise, led to no nontrivial threshold. As we now see, it is clear why existing arguments cannot work: The overcounting of spanning paths, exacerbated by the added disconnected pieces, can lead to the cancellation of terms for general noise such that we cannot argue that the resulting norm still upper-bounds the bad part of the computation. The extra terms, those that do not correspond to logical errors and are hence not in $\sG_\mathrm{bad}$, do not matter for probabilistic noise as all terms sum constructively, These extra terms arise not just from the overcounting (following Reference \cite{FowlerProof}) of the spanning paths to include paths that do not reach the opposite boundary, but also from the fact that added defects in the disconnected pieces can cause the decoder to break up a spanning path such that no logical error results. This is a direct manifestation of the global nature of the surface-code decoding, that the entire set of defects has to be considered; one simply cannot draw conclusions from just a local subset of defects.\\
\medskip

As argued above, merely improving the counting of the disconnected pieces will likely not improve matters. Part of the issue, which may hint at a genuine failure of surface-code quantum computing for general noise, is that the contribution from the disconnected pieces---part of the cost of doing error correction---seems to grow as the volume of the syndrome lattice, while the benefit, namely the removal of errors, only leads to a linear-in-$L$ suppression. Unless one finds a way of arguing that the disconnected contribution is only a unit factor (which appears difficult, as elaborated on in the Methods section), it seems difficult to escape this conclusion that the cost of doing surface-code error correction rapidly outstrips the benefit, and no nontrivial threshold results.\\
\medskip

Again, we emphasize that we are not able to prove definitively that surface-code quantum computing fails under general noise. A proof approach that avoids the pitfalls we pointed out here might still give a nonzero fault-tolerance threshold. We invite the reader to the task, and simply raise the caution that the question of the efficacy of surface-code quantum computing may not be as settled as it may seem at the moment.

\section{Methods}
Here, we provide the technical details of the results discussed in the main text.

\subsection{Probabilistic noise}\label{sec:MethodsProb}
The steps leading to Equation~\eqref{eq:probthreshold} are as follows, starting from the left-hand side of that equation, and evaluated for odd $L=2t+1$ (so $t+1=\lceil L/2\rceil$):
\begin{align}
&\quad \sum_{r=L}^\infty(\textrm{number of length-$r$ paths})P(r)= \sum_{r=L}^\infty \underbrace{4NL}_{\textrm{starting vertex}} \times \underbrace{11^r}_{r \textrm{ other vertices}}\times P(r)\\
&\leq\sum_{r=L}^\infty (2NL)11^rp^{\lceil r/2\rceil}2^r=2NL{\left(\sum_{r=L,r\textrm{ odd}}^\infty 22^rp^{\lceil r/2\rceil}+\sum_{r=L+1,r\textrm{ even}}^\infty 22^rp^{\lceil r/2\rceil}\right)}\nonumber\\
&=2NL{\left(\sum_{s=t}^\infty 22^{2s+1}p^{\lceil (2s+1)/2\rceil}+\sum_{s=t}^\infty 22^{2s+2}p^{\lceil (2s+2)/2\rceil}\right)}\nonumber\\
&=2NL(22)(23)p\sum_{s=t}^\infty (22^2p)^{s}=2NL(22)(23)p\frac{(22^2p)^t}{1-22^2p}=2NL\frac{23}{22}\frac{(22^2p)^{\lceil L/2\rceil}}{1-22^2p}.\nonumber
\end{align}
Note that there are differences between our expression here and the corresponding one in Reference \cite{FowlerProof}, stemming from differences in the handling of the temporal boundaries, and our exact evaluation of the sum in the second line of the above equation (Reference \cite{FowlerProof} only approximated the result). These differences, however, do not affect the conclusions on the accuracy threshold.
\medskip

\subsection{General noise}

Let us first provide the steps towards Equations \eqref{eq:Wprime} and \eqref{eq:W}, before elaborating on how the disconnected pieces can cause problems. Starting from Equation \eqref{eq:sumFP}, we have,
\begin{align}
\Vert\sR\circ\sM\circ\sG'\Vert\leq \Vert\sG'\Vert&\leq \sum_\cP\Vert\cF(\cP)\Vert
\leq \sum_{r=L}^\infty(\textrm{number of such length-$r$ paths})W(r),
\end{align}
where $W(r)$ is the upper bound on the norm of a particular length-$r$ path [analogous to $P(r)$ of Eq.~\eqref{eq:Pr}],
\begin{equation}
W(r)\equiv \sum_{k=\lceil r/2\rceil}^r\binom{r}{k}(2\eta)^k (1-\eta)^{r-k}\Vert\id\Vert\leq (2\eta)^{\lceil r/2\rceil}\sum_{k=0}^r\binom{r}{k}=(2\eta)^{\lceil r/2\rceil}2^r.
\end{equation}
Following the previous argument for probabilistic noise (see Section \ref{sec:MethodsProb}), we then find that, for general noise, we have
\begin{equation}
\Vert\sR\circ\sM\circ\sG'\Vert\leq cNL\frac{(22^2\cdot 2\eta)^{\lceil L/2\rceil}}{1-22^2\cdot 2\eta},
\end{equation}
as given in Equation \eqref{eq:Wprime}.\\
\medskip

Next, let us derive Equation \eqref{eq:W}. Beginning with Equation \eqref{eq:sumFP2},  and noting that $\Vert\cF_a\Vert\leq 2\eta$, we have
\begin{align}
\Vert\sR\circ\sM\circ\sG_\mathrm{bad}\Vert&\leq \sum_\cP\Vert\cF(\cP)\Vert(1+\eta)^{A-|\cP|}\nonumber\\
&\leq \sum_{r=L}^\infty(\textrm{number of length-$r$ paths})W(r)(1+\eta)^{A-r}\nonumber\\
&\leq 2NL(1+\eta)^A\sum_{r=L}^\infty{\left(\frac{22}{1+\eta}\right)}^r(2\eta)^{\lceil r/2\rceil}=cNL(1+\eta)^A\frac{{\left[\frac{22^2\cdot2\eta}{(1+\eta)^2}\right]}^{\lceil L/2\rceil}}{1-\frac{22^2\cdot 2\eta}{(1+\eta)^2}}\,,,
\end{align}
where the last equality follows the same logic as in Section \ref{sec:MethodsProb}, and $c=2(23/22)$ as before.\\
\medskip

Now, let us elaborate on the problems caused by the disconnected pieces, and whether the full $\cN$ can be inserted in our sum of fault paths above for every edge in the disconnected pieces. We first recall the logic leading up to this. We estimated the bad piece, containing the fault paths that lead to logical errors, in the following manner: We first (over)counted all spanning paths---these are fully connected sets of edges that go from one boundary face of the syndrome lattice to the opposite face. Then, we embellished each spanning path with disconnected pieces by adding edges on the syndrome lattice that can have faults or no faults, corresponding to the insertion of $\cF$ or $(1-\eta)\id$, respectively. In $\sG'$, we inserted both pieces, i.e., the full $\cN$, for every edge not on the spanning path, yielding the upper bounds $\WL$ and $\WpL$ depending on when we take the norm. \\
\medskip

\begin{figure}
\centerline{\includegraphics[width=0.95\columnwidth]{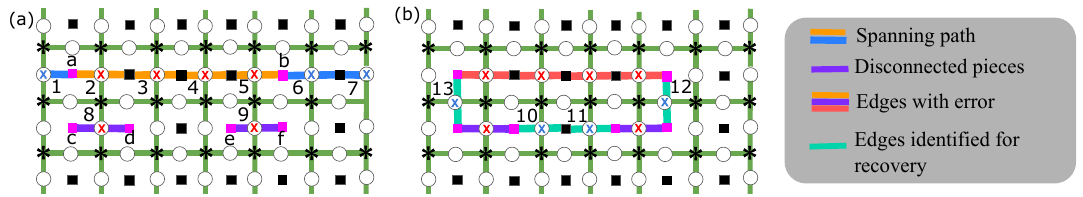}}
\caption{\label{fig:disconnectedPaths}(a) A spanning path as well as disconnected pieces are illustrated. Without the disconnected pieces, the spanning path will be chosen by the decoder (joining the defects to the boundaries), and a logical error results. (b) The additional faults on the disconnected pieces changes the recovery picked by the decoder. No logical error arises in this case.}
\end{figure}

Let us first see how disconnected pieces added to a specified spanning path can result in no logical errors, thus effectively moving the corresponding fault path from the $\sG_\mathrm{bad}$ piece (for the spanning path only) to the $\sG_\mathrm{good}$ piece (for the spanning path together with the disconnected pieces). An illustrative example is given in Figure \ref{fig:disconnectedPaths}, for the $L=7$ surface code. We begin with a specified spanning path [marked in orange and blue in Figure \ref{fig:disconnectedPaths}(a)], corresponding to a fault path with four insertions of $\cF$ as indicated (on edges 2--5). The spanning path results from the MWPM decoder assigning recovery operations at the indicated edges---there are three such recovery edges (edges 1, 6, and 7), and we say that this path has weight 3, counting the number of edges with recovery. The defects manifest at the ancilla locations marked $a$ and $b$. If these faults are the only ones that appear, the spanning path will indeed emerge after the recovery, and a logical error will result. That a four-fault situation can lead to a logical error should come as no surprise since the $L=7$ code guarantees successful correction only if no more than 3 errors occurred.
\\\medskip

Consider, however, adding in some disconnected pieces. Specifically, suppose we have faults at the edges marked $8$ and $9$ in Figure \ref{fig:disconnectedPaths}. Defects will manifest at ancilla locations $c, d, e,$ and $f$. If these were the only faults that appear (i.e., ignoring those on the spanning path), the recovery will correctly remove them by applying recovery operations at edges $8$ and $9$. No logical errors will result.
\\\medskip

The problem arises when we have both the faults on the spanning path and those on the disconnected pieces together. The total weight of the spanning path and the disconnected pieces is $3 +2=5$. This is, however, not the minimum-weight path for the observed defects. Instead, as shown in Figure \ref{fig:disconnectedPaths}(b), there is a weight-4 path connecting the defects on the spanning path with those from the disconnected pieces. This then will be the recovery route determined by the MWPM decoder, and no logical error will result. Adding such disconnected pieces thus lead to the break up of the spanning path, moving the corresponding fault path from $\sG_\mathrm{bad}$ into $\sG_\mathrm{good}$.\\
\medskip

Now, one way of ensuring that we get a large set of terms that contribute only to $\sG_\mathrm{bad}$, i.e., translate into logical errors, is to do the following: For each spanning path, have an ``exclusion zone" around it, such that edges in that exclusion zone not on the spanning path all get only the $\id$ insertion, while edges outside of the exclusion zone get the full $\cN$ insertion. The exclusion zone is one that is large enough (it will grow with size $L^3$) such that any defects that occur outside of it will not be joined by the decoder to any of the defects that occur on the ends of edges on the spanning path. Then, the situation of Figure \ref{fig:disconnectedPaths} cannot occur, and the spanning path remains a spanning one even with defects that arise from the disconnected pieces. The fault paths in this set of terms will contribute only terms with factors of $(1-\eta)^{A'}$ (from the norm of the $\id$ piece), for $A'$ being the number of edges outside the exclusion zone, and a $1$ (from the norm of the full $\cN$) for all the remaining edges. These do not give the $(1+\eta)^A$ factors that caused problems in $\WL$.
\\\medskip

These are, unfortunately, not the only terms that occur in $\sG_\mathrm{bad}$. Faults can occur in the exclusion zone as long as the defects that arise either do not cause the break up of the original spanning path, or that they lead to new spanning paths after the decoder. Going back to the example of Figure \ref{fig:disconnectedPaths}, we observe that if an insertion of a fault $\cF$ occurs only on edge 8 and not on edge 9, the spanning path will remain unbroken even with the extra disconnected piece, and the corresponding fault path remains a part of $\sG_\mathrm{bad}$. It is the simultaneous insertion of $\cF$s at \emph{both} edges 8 and 9 that changes the situation to one with no logical error. This is a clear indication that we cannot have the full $\cN$ inserted at both edges 8 and 9, if we want to be sure that a logical error results. Figuring out the combinatorics of when this happens---and there should be many possibilities since the exclusion zone grows as $L^3$ in size---for all spanning paths seems very challenging.

\bigskip
\medskip
\textbf{Acknowledgements} \par 
This work is supported by a Centre for Quantum Technologies (CQT) Fellowship. CQT is a Research Centre of Excellence funded by the Ministry of Education and the National Research Foundation of Singapore.

\medskip


\end{document}